\documentclass{PoS}

\usepackage{graphicx}
\usepackage{subfigure}
\usepackage{amssymb}
\usepackage{array,multirow}
\usepackage{amsmath}
\usepackage{mathrsfs}
\usepackage{fancyhdr}
\usepackage{verbatim}
\usepackage{color}
\usepackage{bbm}
\usepackage{epstopdf}
\usepackage{etoolbox}

\newcommand{\MSb}{\overline{\mathrm{MS}}}

\newcommand{\oo}{\mathcal{O}}
\newcommand{\lat}{\mathrm{lat}}
\newcommand{\free}{\mathrm{free}}
\newcommand{\massless}{\mathrm{massless}}
\newcommand{\cont}{\mathrm{cont}}

\let\OLDthebibliography\thebibliography
\renewcommand\thebibliography[1]{
  \OLDthebibliography{#1}
  \setlength{\parskip}{0pt}
  \setlength{\itemsep}{0pt plus -0.3ex}
}

\title{{\vspace{-15mm} \normalsize\hfill{\small DESY 16-204}}\\[10mm] 
Step scaling in coordinate space: running of the quark mass}

\ShortTitle{Step scaling in X-space: running of the quark mass}

\author{\speaker{Krzysztof Cichy}\thanks{We thank Pan Kessel for his code to extract the Creutz ratios and discussions at the initial stage of this project. K.C.\ was supported in part by the Deutsche Forschungsgemeinschaft (DFG), project nr. CI 236/1-1 (Sachbeihilfe). Numerical simulations were carried out at the PAX cluster at DESY Zeuthen.}\\
\\ Goethe-Universit\"at Frankfurt am Main, Institut f\"ur Theoretische Physik,
Max-von-Laue-Str. 1, 60438 Frankfurt am Main, Germany       \\
Adam Mickiewicz University, Faculty of Physics,
Umultowska 85, 61-614 Pozna\'n, Poland\\
        E-mail: \email{kcichy@th.physik-uni-frankfurt.de}}

\author{Karl Jansen\\
        John von Neumann Institute for Computing (NIC), DESY, Platanenallee 6, \\
    15738 Zeuthen, Germany\\
        E-mail: \email{Karl.Jansen@desy.de}}

\author{Piotr Korcyl\\
        Institut f\"ur Theoretische Physik, Universit\"at Regensburg, 93040 Regensburg, Germany\\
        E-mail: \email{piotr.korcyl@ur.de}}

\abstract{We perform a benchmark study of the step scaling procedure for the ratios of renormalization constants extracted from position space correlation functions. We work in the quenched approximation and consider the pseudoscalar, scalar, vector and axial vector bilinears. The pseudoscalar/scalar cases allow us to obtain the non-perturbative running of the quark mass over a wide range of energy scales -- from around 17 GeV to below 1.5 GeV -- which agrees well with the 4-loop prediction of continuum perturbation theory.  We find that step scaling is feasible in X-space and we discuss its advantages and potential problems. }

\FullConference{34th annual International Symposium on Lattice Field Theory\\
         24-30 July 2016\\
         University of Southampton, UK}

\begin{document}
\section{Introduction}
\vspace*{-3mm}
One of the non-perturbative renormalization schemes that can be used on the lattice is the coordinate space (X-space) scheme \cite{Martinelli:1997zc,Gimenez:2004me,Cichy:2012is,Tomii:2016xiv}.
In this proceeding, we report shortly on our benchmark study of combining this scheme with the step scaling technique \cite{Luscher:1993gh,Luscher:1991wu,Jansen:1995ck}.
A more complete description of this study is given in Ref.~\cite{Cichy:2016qpu}.
The step scaling technique allows to reliably connect the high energy regime where perturbation theory (PT) can be safely applied to the low energy regime of large volume simulations where hadron matrix elements are evaluated.
It consists in performing several simulations at different lattice spacing and volumes, starting at the coarse, large volume level and repeating the procedure
until the high energy regime of the theory is reached.
In the X-space scheme, the renormalization condition is imposed on correlation functions in coordinate space at very short distances (around 0.01-0.1 fm) compared to the physical extent of the simulated box (an order of magnitude larger). This allows to use infinite volume PT to translate results from the X-space to the $\MSb$ scheme. 
The X-space scheme has certain appealing properties with respect to e.g.\ the RI-MOM scheme, in particular it is gauge invariant and hence no gauge fixing is needed.

Our study is performed in the quenched approximation, hence our results can be compared with continuum perturbation theory setting $N_f=0$.
We choose the standard Wilson plaquette gauge action, using the CHROMA software \cite{Edwards:2004sx} to generate gauge field configurations.
The valence quarks are twisted mass (TM) fermions \cite{Frezzotti:2000nk,Frezzotti:2003ni,Frezzotti:2004wz}, automatically $\mathcal{O}(a)$-improved by tuning  the hopping parameter $\kappa$ to its critical value, such that the PCAC quark mass vanishes \cite{Frezzotti:2003ni,Jansen:2005kk}.

\vspace*{-3mm}
\section{X-space renormalization scheme and step scaling}
\vspace*{-3mm}
Here, we shortly summarize the main ideas of the X-space renormalization scheme.
We consider flavour non-singlet correlation functions of two operators of the form 
\vspace*{-3mm}
\begin{equation}
C_\Gamma(X)\equiv\langle \oo_{\Gamma}(X) \oo_{\Gamma}(0)\rangle, 
\vspace*{-3mm}
\end{equation}
where $\oo_{\Gamma}(X) = \bar{\psi}(X) \Gamma \psi(X),\, \Gamma = 
 \{ 1,\gamma_5, \gamma_{\mu}, \gamma_{\mu} \gamma_5 \}\equiv\{S,P,V,A\}$.
The renormalization condition is imposed in the chiral limit,
\vspace*{-3mm}
\begin{equation}
 \lim_{a\rightarrow 0}\langle \oo^X_{\Gamma}(X) \oo^X_{\Gamma} (0) \rangle
\big|_{X^2=X_0^2} = \langle \oo_{\Gamma}(X_0) \oo_{\Gamma}(0)
\rangle^{\free, \massless}_{\cont},
\vspace*{-3mm}
%\label{eq. condition}
\end{equation}
with four-vectors denoted by capital letters, e.g.~$X=(x,y,z,t)$ and $X^2\equiv x^2+y^2+z^2+t^2$.
Thus, the renormalized operator is $\oo^X_{\Gamma}(X, X_0) = Z^X_{\Gamma}(X_0) \oo_{\Gamma}(X)$,
where $X_0$ is the renormalization point, chosen to satisfy $a\ll \sqrt{X_0^2} \ll \Lambda^{-1}$ to keep discretization effects under control and to ease contact to perturbation theory ($\Lambda$ is a low-energy scale of the order of a few hundred MeV).

We subtract tree-level cut-off effects by computing the ratio of the tree-level lattice and continuum correlators, $\Delta_{\Gamma}(X)$.
The corrected correlation function is $C'_{\Gamma}(X)=C_{\Gamma}(X)/\Delta_{\Gamma}(X)$.
Renormalization constants (RCs), at the scale $\mu=1/\sqrt{X_0^2}$, are given by
\vspace*{-3mm}
\begin{equation}
 Z^X_{\Gamma}(X_0) = \sqrt{\frac{C_{\Gamma}(X_0)_{\cont}^{\free}}{C'_{\Gamma}(X_0)}}
= \sqrt{\frac{C_{\Gamma}(X_0)_{\lat}^{\free}}{C_{\Gamma}(X_0)}}.
\vspace*{-3mm}
\end{equation}
In the end, we want to compare with the running of RCs in the $\MSb$ scheme, which can be accomplished by converting the X-space RCs using 4-loop conversion formulae \cite{Chetyrkin:2010dx}.

To investigate the running of RCs, we use the step scaling method \cite{Luscher:1993gh,Luscher:1991wu,Jansen:1995ck}.
We define the step scaling function as
\vspace*{-3mm}
\begin{equation}
\Sigma^X_{\Gamma}(\mu, 2\mu) = \lim_{a \rightarrow 0} \frac{Z^X_{\Gamma}(2 \mu, a)}{Z^X_{\Gamma}(\mu, a)}, 
\label{eq. sigma}
\vspace*{-3mm}
\end{equation}
with $\mu=1/\sqrt{X_0^2}$ and where the lattice spacing $a$ as an argument of $Z_\Gamma^X$ indicates that it was regularized on the lattice.
Such defined step scaling function has a well-defined continuum limit that we want to find on the lattice.

We perform three steps of the step scaling procedure, allowing us to link non-perturbatively 
the scales between around 1.5 GeV and 17 GeV. For the estimation of $\Sigma^X_{\Gamma}(\mu, 2\mu)$, we need two sets of lattices, with 
spatial extents $L$ and $2L$. 
We look at 3 kinds of points in X-space:  $(x/a,x/a,x/a,x/a)$, $(x/a,x/a,x/a,0)$ and $(x/a,x/a,0,0)$ (where $0$ is at different positions, with respective correlators averaged over them).
We refer to them as points of type IV, III and II, respectively.
We only exclude points of type I, $(x/a,0,0,0)$, known to be affected by very large cut-off effects. 

\vspace*{-3mm}
\section{Ensembles for step scaling}
\vspace*{-3mm}
An important step to carry out the step scaling programme is to have a set of gauge ensembles matched to one another in terms of the physical volume.
Such matching can be done in terms of an effective renormalized coupling, which we define following Creutz \cite{Creutz:1980zw}.
The details of this procedure can be found in Ref.~\cite{Cichy:2016qpu}.
Here, we only show the final result in Tab.~\ref{ensemble}.
Each row of this table contains $\beta$ values such that the physical volumes are matched, e.g. the volume of a $\beta=9.50$, $L/a=64$ lattice is the same as the one of $\beta=7.90$, $L/a=16$. The physical volumes of lattices in the second (third) row are a factor of 2 (4) larger than the ones of the first row.
We note that uncertainty in matching is propagated to the final results for the step scaling function, see below.

\setlength{\tabcolsep}{5pt}
\begin{table}
\begin{center}
\begin{tabular}{|c| c c c c |}
\hline
step & 32/64 & 24/48 & 16/32 & 8/16 \\
\hline
1 & $\beta = 9.50(7)$ & ${\bf \beta = 9.00}$ & $\beta = 8.62(7)$ & $\beta = 7.90(13)$ \\
2 & $\beta = 8.62(7)$ & $\beta = 8.24(6)$ & $\beta = 7.90(13)$ & $\beta = 7.18(2)$  \\
3 & $\beta = 7.90(13)$ & $\beta = 7.56(11)$ & $\beta = 7.18(2)$ & $\beta = 6.61(2)$ \\
\hline
\end{tabular}
\caption{\label{ensemble}Results of our matching procedure. Each entry contains the appropriate value of matching $\beta$ and its error, which is propagated to account for mismatching effects. $\beta=9.00$ is the starting point and hence has no associated uncertainty.
\vspace*{-9mm}}
\end{center}
\end{table}

Knowing the matched values of $\beta$, the last step before computing the appropriate X-space correlation functions is to tune the PCAC mass to zero to achieve maximal twist.
The resulting values of the $\kappa$ parameter are shown in Tab.~\ref{tab:setup}, which is a summary of all our generated ensembles (apart from the ones only used for matching).
We also show the values of the lattice spacings for all ensembles. At $\beta=6.61$, we can express the value of the lattice spacing in terms of the $r_0$ value.
Using the parametrization of Ref.~\cite{Necco:2001xg}, we find $r_0/a=12.76$ for this $\beta$. Together with our chosen value of $r_0=0.48(2)$ fm, we get $a_{\beta=6.61}=0.0376(16)$ fm and using the results of the matching procedure, this provides scale setting for all the ensembles.

\begin{table}[t!]
\begin{center}
\begin{tabular}{cccccccc}
\hline
$\beta$ & $a$ [fm] & $L/a$ & $T/a$ & $\kappa$ & nr of confs & step\\
\hline
9.50 & 0.00235(10) & 64 & 128 & 0.137032& 200 & 40\\
& & 32 & 128&  & 200 & 40\\
9.00 & 0.00314(13) & 48 & 96 & 0.138060& 200 & 40\\
& & 24 & 96 &  & 200 & 40\\
8.62 & 0.00470(20) & 64 & 128 & 0.138976& 200 & 40\\
& & 32 & 128&  & 200 & 40\\
& & 16 & 64&  & 200 & 40\\
8.24 & 0.00627(26) & 48 & 96 & 0.140016& 200 & 40\\
& & 24 & 96 &  & 200 & 40\\
7.90 & 0.00941(39) & 64 & 128 & 0.141173& 200 & 40\\
& & 32 & 128&  & 200 & 40\\
& & 16 & 64&  & 200 & 40\\
& & 8 & 32 &  & 1000 & 40\\
7.56 & 0.01254(52) & 48 & 96 & 0.142512& 200 & 40\\
& & 24 & 96&  & 200 & 40\\
7.18 & 0.01881(78) & 32 & 128 & 0.144324& 200 & 40\\
& & 16 & 64&  & 200 & 40\\
& & 8 & 32 &  & 1000 & 40 \\
6.61 & 0.03763(157) & 16 & 64 & 0.148162 & 200 & 40\\
& & 8 & 32 &  & 1000 & 40\\
\hline
\end{tabular}
\caption{\label{tab:setup}Summary of ensembles used for step scaling: inverse bare coupling $\beta$, lattice spacing in fm (with its uncertainty), lattice size, critical $\kappa$, number of generated configurations, number of heatbath updates between saved configurations.\vspace*{-9mm}}
\end{center}
\end{table}

\vspace*{-3mm}
\section{Procedure}
\vspace*{-3mm}
\label{sec:procedure}
We now summarize our procedure:
\vspace*{-2mm}
\begin{enumerate}
\itemsep-1mm
 \item Compute the relevant correlation functions in X-space, at 3 values of the valence quark mass.
 \item Extrapolate the correlators to the chiral limit, linearly in $a\mu$.
 \item Apply the tree-level correction to the correlators.
 \item Use the chirally extrapolated and tree-level corrected values of correlators to compute the step scaling function.
 \item Extrapolate to the continuum, using the fitting ansatz
 \vspace*{-3mm}
 \begin{equation}
 \quad\Sigma_\Gamma(\mu,2\mu,a)_{\rm corrected}= \Sigma_\Gamma(\mu,2\mu)_{\rm cont} + c_{1} a^2,
 \vspace*{-3mm}
\end{equation}
 \begin{equation}
 \quad\Sigma_\Gamma(\mu,2\mu,a)_{\rm non-corrected}= \Sigma_\Gamma(\mu,2\mu)_{\rm cont} + c_{2} a^2,
 \vspace*{-1mm}
\end{equation}
 i.e.\ a combined fit linear in $a^2$, using only three finest lattice spacings, in total six data points and three fitting parameters, $\Sigma_\Gamma(\mu,2\mu)_{\rm cont}$, $c_{1}$ and $c_{2}$.
 To estimate the systematic uncertainty from the fitting ansatz, we also extend this fit to a quadratic one in $a^2$ and incorporate the coarsest lattice spacings of each step.
 In certain cases, when the cut-off effects in the non-corrected step scaling function are too large, we only consider the corrected one.
 \item Convert from the X-space to the $\MSb$ renormalization scheme.
 \item Calculate systematic uncertainties from non-ideal matching and the uncertainties of $\Lambda_{\MSb}^{(0)}=238(19)$ MeV \cite{Luscher:1993gh} and $r_0=0.48(2)$ MeV. The former are accessed from numerical estimates of the derivative of the step scaling function with respect to $\beta$ and the latter from explicit computations at $\Lambda_{\MSb}^{(0)}=219$ and 257 MeV or $r_0=0.46$ or 0.5 fm.
\end{enumerate}
We note that we also investigated finite volume effects, concluding that they are negligible (well beyond statistical errors) for our choice of $x/a$ values that always satisfy the condition $x/L=1/8$. The only counterexample is when we consider the pair of volumes 8/16 and therefore, we only use the data from it for an estimate of systematics (the preferred fit only uses 3 finest lattice spacings).

\vspace*{-3mm}
\section{Results and discussion}
\vspace*{-3mm}
\label{sec. results}
The results of the step scaling procedure outlined in the previous section are given in Tab.~\ref{tab:step1} for the pseudoscalar (PP) / scalar (SS) correlators and in Tab.~\ref{tab:step2} for the vector (VV) / axial vector (AA) ones, together with the decomposition of all the uncertainties.
This decomposition shows that the most important source of uncertainty is the statistical one, with typically the one from the fitting ansatz or from the matching as the second most important one.
The comparison with 4-loop $N_f=0$ continuum perturbation theory \cite{Chetyrkin:1997dh,Vermaseren:1997fq}, or the exact value of 1 for the VV/AA case, is given in Tab.~\ref{tab:step1a}.
In most cases, the agreement between our lattice results extrapolated to the continuum limit and PT is satisfactory.
However, we observe some regularities depending on the type of points that we consider, e.g.\ for the SS case, points of type II(IV) tend to lie above(below) the PT result and points of type III tend to agree best with PT.
This suggests systematically different cut-off effects for these kinds of points, which indicates that the X-space approach could be improved by better understanding of hypercubic artefacts.

\setlength{\tabcolsep}{2pt}
\begin{table}[t!]
\begin{center}
\begin{footnotesize}
\begin{tabular}{ccccc}
\hline
$\mu$ & $2\mu$ & point & $\Sigma_P^{\MSb}(\mu,2\mu)$  & $\Sigma_S^{\MSb}(\mu,2\mu)$ \\
${\rm [GeV]}$ & ${\rm [GeV]}$ & type & lattice & lattice \\
\hline
1.478 & 2.956 & IV &  1.0995(104)(66)(33)(13)(37) & 1.1134(121)(56)(37)(13)(37)\\ 
1.706 & 3.413 & III &  1.1027(91)(19)(36)(10)(29) & 1.1210(115)(6)(41)(11)(29)\\
2.090 & 4.180 & II &  1.1012(101)(33)(49)(8)(23) & 1.1337(140)(13)(52)(8)(23)\\
2.956 & 5.911 & IV &  1.0787(81)(31)(21)(4)(16) & 1.0856(90)(27)(21)(5)(16)\\
3.413 & 6.826 & III &  1.0743(72)(18)(19)(4)(14) & 1.0846(90)(14)(18)(4)(14)\\
4.180 & 8.360 & II &  1.0691(78)(22)(23)(3)(12) & 1.0961(109)(1)(22)(3)(12)\\
5.911 & 11.822 & IV &  1.0721(57)(22)(3)(2)(9) & 1.0728(71)(30)(3)(2)(9)\\
6.826 & 13.651 & III &  1.0736(57)(35)(4)(2)(9) & 1.0802(73)(8)(5)(2)(9)\\
8.360 & 16.719 & II &  1.0571(65)(24)(5)(2)(8) & 1.0755(91)(1)(7)(1)(8)\\
\hline
\end{tabular}
\end{footnotesize}
\caption{\label{tab:step1}Step scaling function $\Sigma_{P/S}(\mu,2\mu)$, from PP/SS correlators. We show the scale change, the type of points used  and our continuum-extrapolated result translated to the $\MSb$ scheme. The uncertainties are: statistical, from the fitting ansatz, from matching, from $\Lambda_{\MSb}^{(0)}$ and from $r_0$.\vspace*{-5mm}}
\end{center}
\end{table}

\setlength{\tabcolsep}{4pt}
\begin{table}[t!]
\begin{center}
\begin{footnotesize}
\begin{tabular}{ccccc}
\hline
$\mu$ & $2\mu$ & point & $\Sigma_V^{\MSb}(\mu,2\mu)$  & $\Sigma_A^{\MSb}(\mu,2\mu)$ \\
${\rm [GeV]}$ & ${\rm [GeV]}$ & type & lattice & lattice \\
\hline
1.478 & 2.956 & IV & 0.9918(103)(1)(15)(2)(1) & 0.9931(86)(30)(9)(2)(1)\\
1.706 & 3.413 & III & 0.9968(108)(6)(22)(2)(1) & 0.9987(87)(20)(16)(2)(1)\\
2.090 & 4.180 & II & 1.0127(99)(15)(24)(1)(1) & 0.9683(69)(61)(23)(1)(1)\\
2.956 & 5.911 & IV &  1.0061(87)(10)(6)(1)(1) & 1.0039(70)(12)(6)(1)(1)\\
3.413 & 6.826 & III & 1.0122(92)(2)(10)(1)(0) & 1.0092(74)(19)(8)(1)(0)\\
4.180 & 8.360 & II & 1.0273(85)(22)(12)(1)(0) & 0.9848(59)(60)(8)(1)(0)\\
5.911 & 11.822 & IV & 1.0017(69)(12)(0)(1)(0) & 1.0009(58)(6)(2)(1)(0)\\
6.826 & 13.651 & III & 1.0103(77)(12)(2)(1)(0) & 1.0085(63)(27)(3)(1)(0)\\
8.360 & 16.719 & II &  1.0125(73)(14)(8)(1)(0) & 0.9823(52)(53)(1)(1)(0)\\
\hline
\end{tabular}
\end{footnotesize}
\caption{\label{tab:step2}Step scaling function $\Sigma_{V/A}(\mu,2\mu)$, from AA/VV correlators. We show the scale change, the type of points used  and our continuum-extrapolated result translated to the $\MSb$ scheme. The uncertainties are: statistical, from the fitting ansatz, from matching, from $\Lambda_{\MSb}^{(0)}$ and from $r_0$.\vspace*{-9mm}
}\end{center}
\end{table}

\setlength{\tabcolsep}{2pt}
\begin{table}[t!]
\begin{center}
\begin{footnotesize}
\begin{tabular}{cc|ccc|ccc}
\hline
 $\mu$ & $2\mu$ & $\Sigma_{P/S}^{\MSb}(\mu,2\mu)$ & $\Sigma_P^{\MSb}(\mu,2\mu)$ & $\Sigma_S^{\MSb}(\mu,2\mu)$ & $\Sigma_{V/A}^{\MSb}(\mu,2\mu)$ & $\Sigma_V^{\MSb}(\mu,2\mu)$ & $\Sigma_A^{\MSb}(\mu,2\mu)$\\
${\rm [GeV]}$ & ${\rm [GeV]}$ & 4-loop PT & lattice & lattice & exact & lattice & lattice\\
\hline
1.478 & 2.956 & 1.1318(68) & 1.0995(133) & 1.1134(144) & 1.0 & 0.9918(104) & 0.9931(92)\\
1.706 & 3.413 & 1.1206(56) & 1.1027(104) & 1.1210(126) & 1.0 & 0.9968(110) & 0.9987(91)\\
2.090 & 4.180 & 1.1080(44) & 1.1012(120) & 1.1337(152) & 1.0 & 1.0127(103) & 0.9683(95)\\
2.956 & 5.911 & 1.0919(31) & 1.0787(91) & 1.0856(98) & 1.0 & 1.0061(88) & 1.0039(71)\\
3.413 & 6.826 & 1.0866(27) & 1.0743(78) & 1.0846(94) & 1.0 & 1.0122(93) & 1.0092(77)\\
4.180 & 8.360 & 1.0802(23) & 1.0691(85) & 1.0961(112) & 1.0 & 1.0273(89) & 0.9848(85)\\
5.911 & 11.822 & 1.0713(18) & 1.0721(62) & 1.0728(78) & 1.0 & 1.0017(70) & 1.0009(58)\\
6.826 & 13.651 & 1.0682(16) & 1.0736(68) & 1.0802(74) & 1.0 & 1.0103(78) & 1.0085(69)\\
8.360 & 16.719 & 1.0643(15) & 1.0571(70) & 1.0755(92) & 1.0 & 1.0125(75) & 0.9823(74)\\
\hline
\end{tabular}
\end{footnotesize}
\caption{\label{tab:step1a}Comparison of lattice results for the step scaling function $\Sigma_{P/S/V/A}(\mu,2\mu)$, with continuum 4-loop perturbation theory or the exact value of 1 for VV/AA. The errors of the lattice result were combined in quadrature.\vspace*{-9mm}}
\end{center}
\end{table}

\begin{figure}[t!]
\begin{center}
\includegraphics[width=0.42\textwidth,angle=270]{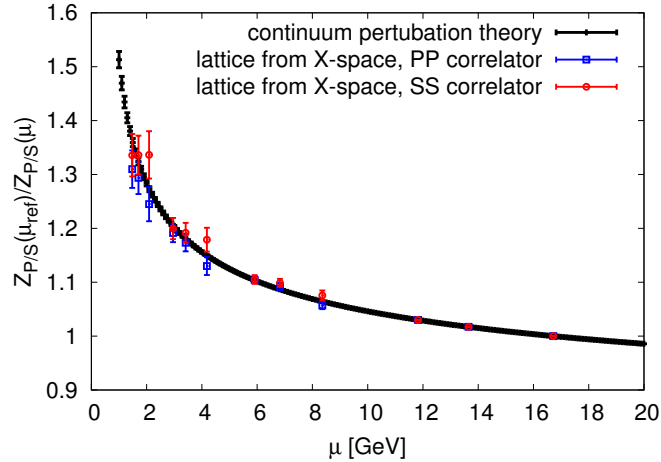}
\caption{\label{fig:run1}Running of step scaling function in the PP/SS case (running of the quark mass). The black symbols correspond to continuum perturbation theory (with an uncertainty related to the uncertainty of $\Lambda_{\MSb}^{(0)}$). The reference scale is $\mu_{\rm ref}=16.719$ GeV. The three rightmost points are the starting points for our step scaling procedure and hence have no errors. To the left of these, there are three groups of three points, corresponding to the three step scaling steps and the three types of points. The rightmost point of each group corresponds to points of type II, the middle one to type III and the leftmost to type IV. \vspace*{-9mm}}
\end{center}
\end{figure}

\begin{figure}[t!]
\begin{center}
\includegraphics[width=0.42\textwidth,angle=270]{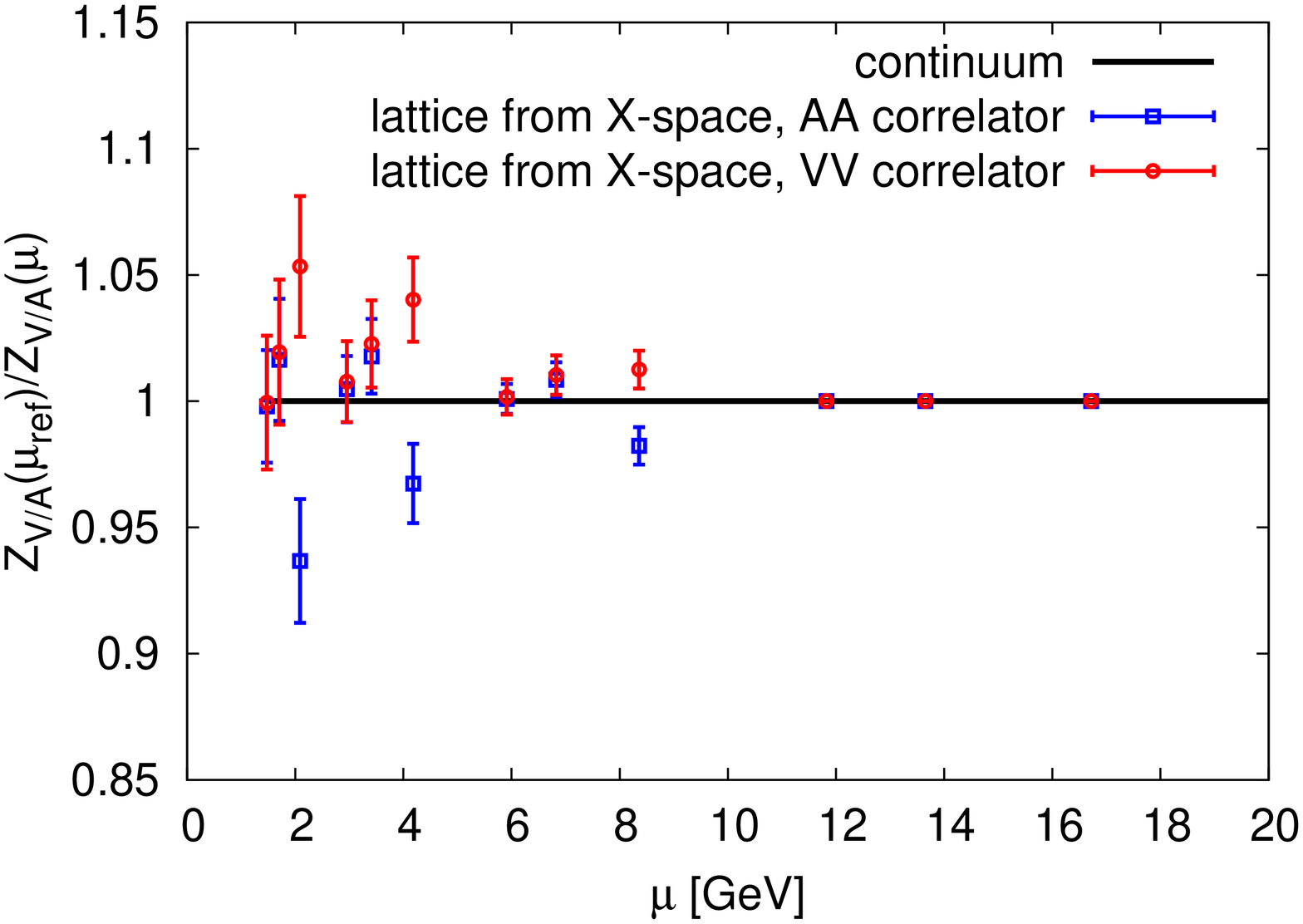}
\caption{\label{fig:run2}Running of step scaling function in the VV/AA case. The black solid line is the exact result of 1.0. The reference scale is $\mu_{\rm ref}=16.719$ GeV. The three rightmost points are the starting points for our step scaling procedure and hence have no errors. To the left of these, there are three groups of three points, corresponding to the three steps of step scaling and to the three types of points. The rightmost point of each group corresponds to points of type II, the middle one to type III and the leftmost to type IV.
\vspace*{-7mm}}
\end{center}
\end{figure}

The full running of the step scaling function is shown in Fig.~\ref{fig:run1} for the PP/SS case and in Fig.~\ref{fig:run2} for VV/AA.
The running obtained from the SS correlator is well reproduced, particularly for points of type IV/III.
Points of type II are approx.\ 1-$\sigma$ above the continuum curve.
In the PP running, we observe a possible tendency that the step scaling function is below its continuum value.
This happens for all types of points, but the result is always within 1-$\sigma$ of the continuum result even in the last step.
Although this can only be a statistical fluctuation, this observation should be investigated further.
The computation of ``running'' for the VV/AA case serves as a cross-check of the method, since the continuum value of 1 is known exactly.
Again, we observe good agreement with this result for points of type III and IV.
However, for type II, deviations from 1 are increasing when decreasing the energy scale, with a final 2-2.5-$\sigma$ discrepancy in the last step, above (below) 1 for VV (AA).
This systematic effect strenghtens the conclusion that hypercubic artefacts should be understood more.
This can be done using techniques similar to those of Ref.~\cite{deSoto:2007ht} (fitting of hypercubic artefacts) or by computing the leading order corrections of $\mathcal{O}(a^2g^2)$ in lattice perturbation theory.

In conclusion, we investigated, for the first time, the step scaling technique using the coordinate space renormalization scheme.
We performed a feasibility study in the quenched approximation, thus reducing the computational cost to a tractable level.
Carrying out three steps of the step scaling, we evaluated the running of the renormalization constants of the pseudoscalar and scalar densities, as well as of the vector and axial vector currents, finding rather good agreement with continuum perturbation theory.
In this way, we demonstrated that the X-space method can provide reliable results.
However, we also conclude that better understanding of hypercubic artefacts will be very important for futher advancement of the method, in particular for its reliable use with dynamical fermions.

\bibliographystyle{JHEP-notitle}
\vspace*{-3mm}
\providecommand{\href}[2]{#2}\begingroup\raggedright\endgroup
\end{document}